\documentclass[12pt,nonatbib]{elsarticle}    
 
\makeatletter
\def\ps@pprintTitle{%
  \let\@oddhead\@empty
  \let\@evenhead\@empty
  \let\@evenfoot\@oddfoot
}
\makeatother

\usepackage{amssymb}

\usepackage[numbers,sort&compress]{natbib}
\usepackage{amsmath}
\usepackage{comment}
\usepackage{makecell}
\usepackage{float}
\usepackage[left=2.54 cm,right=2.54 cm,top=2 cm,bottom=2 cm]{geometry}
\usepackage{ulem}
\usepackage[dvipsnames]{xcolor}
\usepackage{multirow}
\usepackage{caption}
\usepackage{gensymb}
\usepackage{soul}
\usepackage{algorithmic}

\begin{document}
\sloppy
\captionsetup{belowskip=10pt}

\begin{frontmatter}

\title{Effects of Edge Atoms and Channel Width on Charge Storage in Nanoporous Carbon Supercapacitors}

\author[1]{Koushik Sarkar}
\ead{koushik1564@gmail.com}
             
\author[1]{Muhammad Anisuzzaman Talukder\corref{cor1}}
\ead{anis@eee.buet.ac.bd}
\cortext[cor1]{Corresponding author}
\affiliation[1]{organization={Department of Electrical and Electronic Engineering, Bangladesh University of Engineering and Technology},
             city={Dhaka},
             postcode={1205},
             country={Bangladesh}}


\begin{abstract}
The amorphous structure of nanoporous carbon electrodes in supercapacitors complicates the establishment of clear links between electrode geometry and capacitance. In this work, we examine how specific structural features govern charge storage and explain capacitance variations among carbide-derived carbon (CDC) electrodes. Two key methodological and mechanistic advances are introduced. First, we propose a physically motivated, electric-field-based definition of the ion sphere of influence, which avoids the ambiguity of radial distribution function-based cutoffs and enables a more reliable characterization of local ion--electrode interactions. Second, we introduce an ion-resolved channel-width descriptor that quantifies pore accessibility by identifying the narrowest passage a counter-ion must traverse to enter a pore, directly linking accessibility to ion decoordination. Using atomistic molecular dynamics simulations of supercapacitors comprising realistic CDC electrodes and a room-temperature ionic liquid electrolyte under applied potentials, we show that carbon atoms located at the edges of graphitic sheets consistently accumulate higher charge than basal-plane atoms across all electrode types. The fraction of edge atoms increases with structural disorder and correlates with enhanced capacitance, in agreement with recent experimental findings linking disorder to charge storage. Furthermore, analysis of pore size and channel width reveals that ion decoordination is governed primarily by channel width rather than pore size alone. Together, these results establish edge atom concentration and pore channel width as decisive structural descriptors controlling charge accumulation in nanoporous carbon supercapacitors, providing clear design guidelines for optimizing electrode architectures.

\end{abstract}

\end{frontmatter}


\section{Introduction} 
\label{introduction}
The transition from fossil fuels to renewable energy sources is essential to mitigate environmental damage and address the depletion of natural resources like oil, coal, and gas \cite{zhao2021electrochemical}. Efficient electrochemical energy storage (EES) devices are crucial for storing energy generated from renewable sources and for stabilizing power supply \cite{yang2011electrochemical}. Currently, lithium-ion batteries are the leading EES technology used in various industrial applications \cite{deng2015li}. However, supercapacitors (SCs) offer superior cycling stability and faster charging and discharging rates compared to traditional batteries \cite{conway2013electrochemical}. The charge storage mechanism in batteries relies on redox reactions and the intercalation and deintercalation of ions, resulting in slower charge transfer kinetics and material degradation over time \cite{zubi2018lithium}. In contrast, electrochemical double-layer supercapacitors (SCs) store charge by adsorbing ions at the electrode-electrolyte interface \cite{scibioh2020materials}, allowing for rapid charge transfer and enhanced material stability. 

Despite their advantages over traditional EES devices, SCs have a major drawback: their energy density is relatively low compared to batteries \cite{christen2000theory}. Recent efforts have focused on enhancing the SC energy density without sacrificing power density \cite{raza2018recent}. A promising strategy involves using activated carbon for electrodes and room temperature ionic liquid (RTIL) as the electrolyte. Activated carbon materials feature a high surface area, well-defined pore size distribution (PSD), alongside high conductivity and low production costs. RTILs enable a wide operating potential window of 3.5 to 4 volts \cite{zhong2015review}. These properties of the electrodes and electrolytes collectively contribute to enhanced performance, described by the energy formula \(E = \frac{1}{2}CV^2\), where \(C\) is the capacitance and \(V\) is the applied voltage. 

The anomalous relationship between capacitance and pore size in porous carbon SCs has attracted significant interest \cite{chmiola2006anomalous}. The initial hypothesis suggested that the most efficient adsorption of ions would occur if the pore size perfectly matched the ion size, minimizing the void space within the pore \cite{largeot2008relation}. However, as pore size increased, the distance between the adsorbed ion and the pore walls also increased, leading to a reduction in capacitance. Subsequent studies using different electrodes synthesized through various methods did not support this initial hypothesis linking pore size to capacitance. For example, a study involving 28 different nanoporous carbon electrodes using tetraethylammonium (TEA$+$) tetrafluoroborate (BF$_4^-$) in an acetonitrile electrolyte found that the specific capacitance remained relatively constant across average electrode pore sizes ranging from 0.7 to 15 nm \cite{centeno2011capacitance}. This discrepancy suggests that variations in conclusions from different studies may arise from the differing characterization methods used on nanoporous electrodes, which can yield varying surface area values \cite{stoeckli2013optimization}.

Rigorous simulation studies have been performed to clarify the charging mechanism and correlate electrode structural features with capacitance \cite{zhang20232023}. However, investigating SCs using porous carbon presents challenges due to the amorphous nature of the electrodes and difficulties in accurately modeling these structures. Most studies rely on an atomistic model of titanium carbide-derived carbon (CDC) developed by Palmer et al.~\cite{palmer2010modeling}. Notably, the radial distribution function (RDF) of these models shows poor agreement with the X-ray diffraction patterns of experimental samples \cite{de2017structural}, indicating discrepancies in the short-range structural features.
 
The influence of porous structures on the local environments of adsorbed ions was examined by defining a sphere of influence for counter-ions \cite{merlet2013highly}. It was found that desolvation and local electrode charge associated with counter-ions positively correlated with the degree of confinement (DoC). A subsequent study indicated that higher capacitance was achieved in electrodes where counter-ions exhibited a higher DoC \cite{lahrar2021carbon}. In contrast, another investigation of zeolite-templated carbon (ZTC) electrodes revealed a slight negative correlation (Pearson coefficient $= -0.387$) between gravimetric capacitance and averaged DoC values \cite{liu2019carbons}. Overall, this study failed to find a strong correlation between equilibrium capacitance and any geometric descriptor of the electrodes. Another study demonstrated that pore accessibility notably impacts ion percolation \cite{vasilyev2019connections}. Even if a pore is large enough to host an ion, a narrow channel width can render it inaccessible, significantly reducing capacitance \cite{lahrar2021carbon}. The influence of channel width on the local environment of specific adsorbed ions, however, was not fully addressed.

While informative, these studies did not establish a clear interdependence between capacitance and the structural features of SCs. One study explored variations in capacitance by examining ion migration kinetics within charged porous structures \cite{pak2016molecular}. SCs with three electrodes of varying average pore sizes underwent molecular dynamics (MD) voltammetry simulations. The electrode that supported more efficient ion migration during charge/discharge exhibited greater capacitance. A new geometric descriptor, the pore size factor (PSF), was introduced to quantify the property influencing ion migration, showing a correlation between PSF and capacitance. Nonetheless, experimental data revealed significant variation around the theoretical trend, suggesting the influence of unidentified factors. Additionally, recent advancements in experimental characterizations \cite{prehal2017quantification, zhao2023cp} and improved atomistic models of porous carbon \cite{sarkar2024structurally, de2017structural} highlight the need for further simulation investigations that incorporate insights from these novel characterizations.

Previously, we investigated the influence of nanoporous electrodes with varying PSDs on the confinement and decoordination of adsorbed ions \cite{sarkar2025influence}. We found that electrodes with smaller average pore sizes confined ions more effectively, albeit with reduced overall ion admission. This competition led to maximum capacitance occurring in electrodes with intermediate average pore sizes. The capacitance trends observed were primarily attributed to the ion properties, particularly their coordination numbers and total number of adsorbed ions. In this study, we advanced our findings by exploring the relationship between charge accumulation and the geometric features of the electrode, specifically channel width and disorder degree. We analyzed four distinct atomistic electrode models synthesized at varying temperatures, each displaying unique PSDs. A novel approach was developed to define the sphere of influence of ions based on the electric field. The maximum capacitance achieved was 100.6 F/g for the CDC-1500 electrode, while the CDC-2500 electrode exhibited a minimum capacitance of 72.3 F/g, representing a 28\% decrease compared to the CDC-1500, closely aligning with the experimental value of 31\% \cite{chmiola2006anomalous}.

We calculated the number of edge and basal plane atoms in different electrode types, revealing that disordered structures had a higher proportion of edge atoms. The count of edge atoms, indicative of disorder, increased at lower synthesis temperatures. Notably, edge atoms were found to be more polarized than those on the basal planes. Under a 2-V applied potential difference, edge atoms in the CDC-1000 electrode accumulated approximately $\sim$0.5$e$ more charge than basal plane atoms, supporting experimental data that demonstrate higher specific capacitance in more disordered structures \cite{liu2024structural}. Furthermore, we proposed an efficient method for calculating pore channel width that accommodates counter-ions. The most decoordinated ions, characterized by a coordination number of zero, were primarily found in pores with channel widths $<4.67$ \AA. Additionally, the trend of increasing average channel width with rising coordination number of counter-ions highlighted the significant role of channel width on ion coordination.

%
\section{Computational Details} 
\label{computational_details}

%
\subsection{Supercapacitor structure}

\begin{figure}[H]
    \centering
    \includegraphics[width=0.5\textwidth]{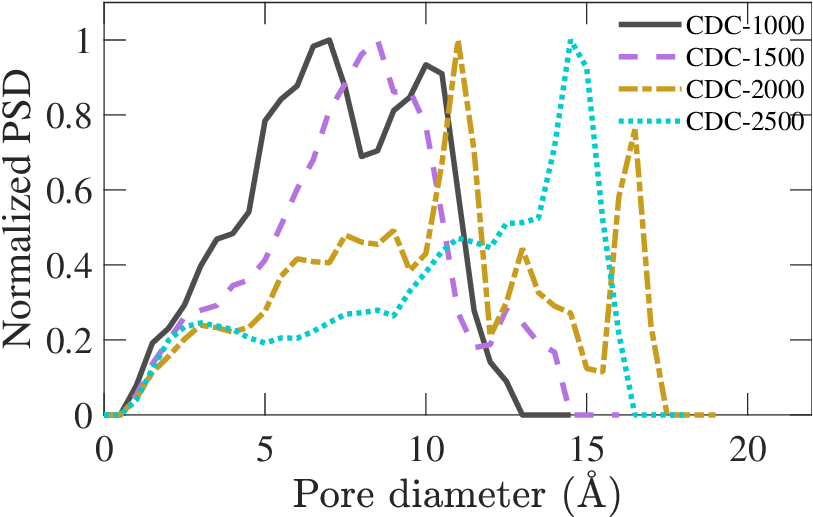}
    \caption{Normalized pore size distribution (PSD) of the atomistic model of carbide-derived carbon (CDC) materials incorporated in the supercapacitor (SC) as electrodes.}
    \label{fig:norm_PSD}
\end{figure}
%

%
\begin{table}[]
    \centering
    \caption{Structural parameters of supercapacitor devices. The accessible surface area and volume were calculated with a probe size of 1.7 \AA.}
    \resizebox{1\textwidth}{!}
    {
    \begin{tabular}{c c c c c}
    \hline
    Electrode type & \makecell{Average \\ pore size (\AA)} & \makecell{Accessible \\ surface area (m$^2$/g)} & \makecell{Accessible \\ volume (cm$^{3}$/g)} & \makecell{Number of electrolyte \\ ion pairs} \\
    \hline

    CDC-1000 & 7.18 & 1522.5 & 0.2 & 640 \\

    CDC-1500 & 7.76 & 1643.5 & 0.23 & 670 \\

    CDC-2000 & 9.92 & 1334.2 & 0.27 & 730 \\

    CDC-2500 & 10.56 & 1181.6 & 0.28 & 720 \\

    \hline
    \end{tabular}
    }
    \label{tab:SC_str_params}
\end{table}
%
 
The atomistic model of CDC materials was created using an annealing methodology, demonstrating good agreement with the structural properties of experimentally synthesized TiC-derived carbon samples \cite{sarkar2024structurally}. The MD simulation temperatures of 1000, 1500, 2000, and 2500 K correspond to experimental synthesis temperatures of 400, 600, 800, and 1000$^{\circ}$ C, respectively. For clarity, the electrodes are designated by their simulation temperatures: CDC-1000, CDC-1500, CDC-2000, and CDC-2500. As shown in Fig.~\ref{fig:norm_PSD}, the PSD profiles shift rightward with increasing synthesis temperature. Table \ref{tab:SC_str_params} indicates an increasing trend in average pore size with temperature. The structure synthesized for the SC simulation had a limited dimension of approximately 4.4 nm, restricting large pore formation beyond 17 \AA. Consequently, the average pore sizes were smaller than those found in larger structures ($\sim$10 nm) \cite{sarkar2024structurally}, although the general trend with temperature remained consistent.

\begin{figure}[H]
    \centering
    \includegraphics[width=0.4\textwidth]{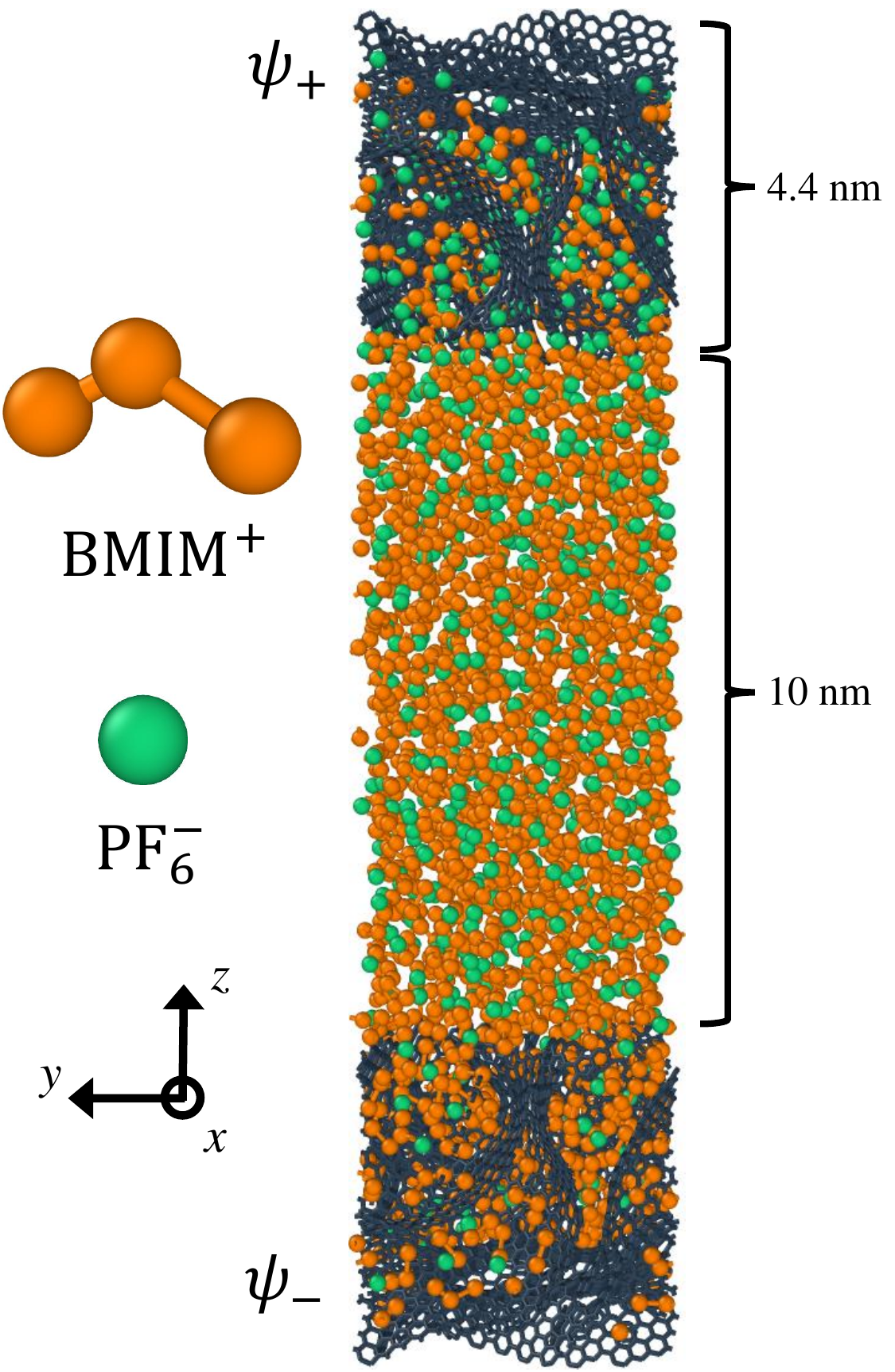}
    \caption{Supercapacitor (SC) structure consisting of room temperature ionic liquid (RTIL), BMIM$^+$PF$_6^-$ and nanoporous carbon electrodes, with the top and bottom electrodes held at a positive ($\psi_+$) and negative ($\psi_-$) potential, respectively. The orange and green spheres represent the BMIM$^+$ and PF$_6^-$ ions, respectively. The nanoporous carbon electrodes are shown by the network formed by the atomic bonds.}
    \label{fig:SC_sch}
\end{figure}

The CDC structures were generated using three-dimensional periodic boundary conditions (PBC), and any dangling bonds from the $z$-surfaces, as shown in Fig.~\ref{fig:SC_sch}, were eliminated before integration into the SC model. For computational efficiency, a coarse-grained model of 1-butyl-3-methylimidazolium hexafluorophosphate (BMIM$^+$PF$_6^-$) RTIL was utilized \cite{roy2010improved}. This model features a single-site anion and a three-site rigid cation. The SC structure was formulated by positioning two cubic CDC structures, generated at the same temperature, 10 nm apart along the $z$-direction, and filling the intervening space with electrolyte ion pairs, as illustrated in Fig.~\ref{fig:SC_sch}. The number of ion pairs was adjusted to achieve a bulk electrolyte density close to the experimental value after equilibrating the structure. Relevant structural parameters for the SCs with different electrode types are summarized in Table~\ref{tab:SC_str_params}.
%
%

\subsection{Molecular dynamics methodology}
The MD simulations were conducted using the LAMMPS software package \cite{plimpton1995fast}. Throughout the simulations of the supercapacitor, the positions of the electrode atoms were constrained. The interactions among the species were modeled using the Coulombic interaction and the 12-6 Lennard-Jones potential. Long-range electrostatic interactions were calculated using the particle-particle/particle-mesh (PPPM) method \cite{pollock1996comments}. 

Periodic boundary conditions were applied in the $x$ and $y$ dimensions, while a fixed boundary condition was implemented in the $z$ direction, with a slab factor of 3.0 for dipole correction. The initial structures of the SC were energy-minimized using the conjugate gradient algorithm, with an energy and force tolerance of $10^{-8}$ and $10^{-8}$ eV/\AA, respectively. The structures were then equilibrated at 400 K for 10 ns, with temperature control achieved through the Nose-Hoover thermostat, which had a relaxation time constant of 10 ps. During this initial stage, the charge of the electrode atoms remained constant. 

In the subsequent equilibration phase, both electrodes of the supercapacitor were held at null potential for 5 ns using the constant potential method (CPM) \cite{siepmann1995influence, pollock1996comments, ahrens2022electrode}, allowing the charge on the electrode atoms to fluctuate and reach a steady state. After charge and temperature equilibration, a constant potential difference was applied between the top and bottom electrodes of the supercapacitor. This phase continued until the total charge on the electrodes approached saturation. The duration for this stage varied from 30 to 100 ns, depending on the type of electrode and the applied potential. We considered five different applied potentials of 2, 2.5, 3, 3.5, and 4 V for simulations. The atomic positions and charges were recorded every 5 ps during the last 5 ns of the simulation to calculate the averaged parameters.

%
\subsection{Electric field calculation}

Previous studies have defined the distance corresponding to the first minimum of the ion-carbon RDF profile as the radius of the sphere of influence or coordination shell of ions \cite{merlet2013highly, lahrar2021carbon}. However, it has been reported that some materials do not exhibit a clear minimum in the RDF profile \cite{liu2019carbons}. Therefore, a more refined approach to calculating this radius is necessary. 
The sphere of influence is employed to quantify the charge induced on electrode atoms by individual electrolyte ions. Consequently, we explored the electric field distribution, magnitude, and direction of the Coulombic force induced on electrode atoms to determine the radius of the sphere of influence. Our analysis revealed that defining the radius based on the electric field analysis yielded consistent values across different types of electrodes and various applied potentials. The electric field on a carbon atom is calculated by
\begin{equation}
    \label{eq:Efield}
    \vec{E}_i = \frac{1}{4\pi{\epsilon}_0}\sum_{k\in N}\frac{{q_k}}{ |\vec{r}_{ik}|^3}\vec{r}_{ik},
\end{equation}
where $q_k$ and $\vec{r_{ik}}$ are the atomic charge of the $k$-th ion and the distance from the $i$-th electrode atom to the center of mass of the ion. The ions in this equation include both electrolyte ions and polarized carbon atoms.

\begin{figure}[H]
    \centering
    \includegraphics[width=1\textwidth]{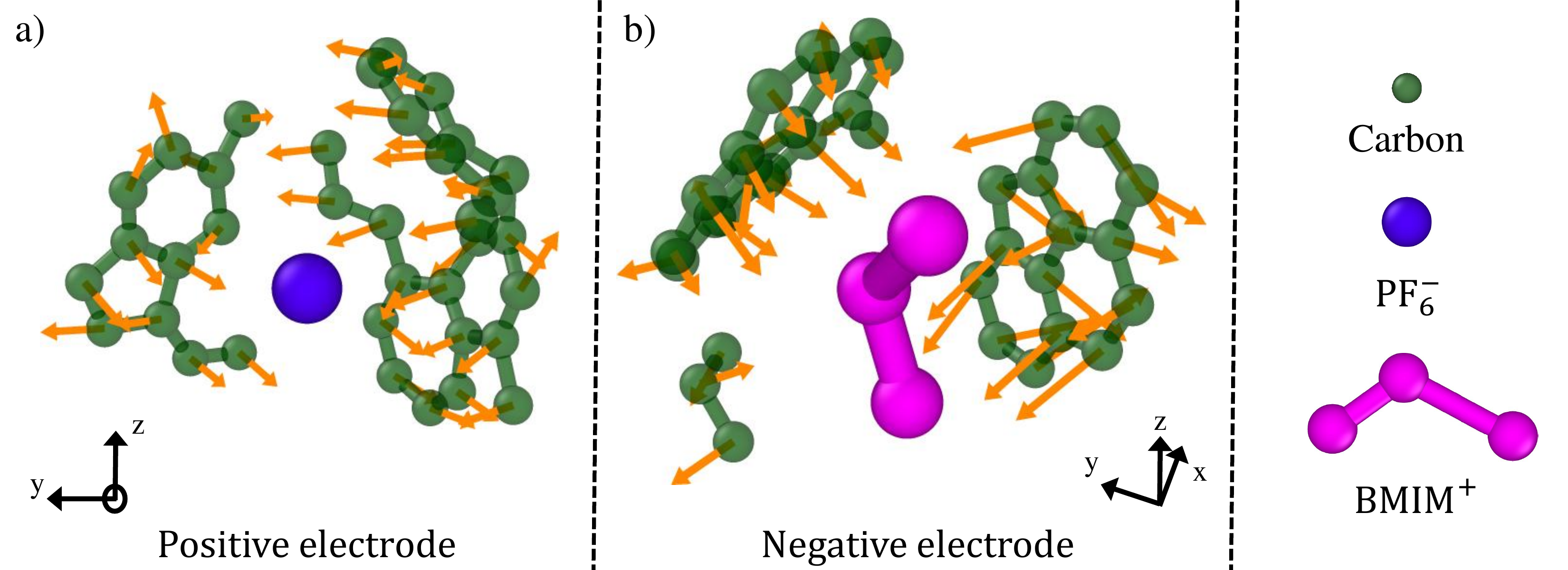}
    \caption{Schematic diagrams of cross-sectional views of (a) positive, and (b) negative electrode with electrode atoms surrounding a counter-ion showing the Coulomb force unit vectors denoted by orange arrows. The purple, magenta, and green spheres represent the PF$^-_6$, BMIM$^+$, and carbon atoms, respectively.}
    \label{fig:sch_efield}
\end{figure}

Two representative cases illustrating the Coulomb force experienced by counter-ions in positive and negative electrodes are shown in Fig.~\ref{fig:sch_efield}. The force unit vectors for most carbon atoms are closely aligned with the direction of the counter-ion. The slight misalignment of the force unit vectors from the counter-ion direction can be attributed to the influence of surrounding polarized carbon atoms. In some instances, the force unit vector even points away from the counter-ion, because either these specific atoms were influenced by another counter-ion or their surrounding carbon atoms were strongly polarized, thereby overcompensating for the impact of counter-ions.
%
%

The electrode atoms that are polarized by an electrolyte ion were identified by analyzing the profile of the 
scalar component of the Coulomb force along the vector that stretches from the electrode atom to the electrolyte ion. The Coulomb force component ($F^{||}_{ij}$) for the $i$-th electrode atom due to the $j$-th electrolyte ion was calculated using the following formulas
\begin{subequations}
\begin{align}
    \vec{F}_{i} &= q_i\vec{E}_i \label{eq:cou_force},\\
%
    F^{||}_{ij} &= |F_i| \frac{\vec{F_i} \cdot \vec{r}_{ij}}{ | \vec{F}_i | | \vec{r}_{ij} | }\label{eq:force_comp},
\end{align}
\end{subequations}
%
where $F_i$ and $\vec{r}_{ij}$ are the Coulomb force experienced by an electrode atom and the distance vector from the electrode atom to the center of mass of the electrolyte ion.
%
\subsection{Channel width calculation}
Inter-pore connectivity is a crucial property of porous materials that affects ion percolation. For ion adsorption within the pores of an electrode, the pores must be sufficiently large to accommodate the ions and accessible from the electrode-bulk electrolyte interface. To characterize inter-pore connectivity, we calculated the channel width associated with each pore containing ions. This methodology involves uniformly distributing grid points throughout the three-dimensional domain of a CDC electrode structure, with a grid spacing set to 1 \AA. Given the maximum dimensions of the electrolyte ions, BMIM$^+$ (9.6 \AA) and PF$_6^-$ (5.06 \AA), the chosen grid spacing is sufficiently small to prevent the assignment of multiple ions to a single grid point. While smaller grid spacings could yield more accurate estimates of channel widths, they would also necessitate increased memory usage. For each grid point, we calculate the radius of the largest sphere that can fit inside the electrode without overlapping any electrode atoms. Each grid point is then assigned to a channel, which is defined as a sequence of grid points connecting the electrode-electrolyte interface to that point. The channel width is determined by the smallest radius of the fitted sphere associated with the grid points forming the channel. If a grid point is accessible by multiple channels, the channel width at that point is defined as the largest width among the channels.

In theory, we could determine the channel width for all grid points by starting at a grid point on the interface and moving to adjacent points in various directions, tracing out all possible channels. However, the number of potential paths increases rapidly with finer grid spacing, making this method computationally infeasible. To address this, we developed a computationally efficient method for determining the channel width based on two core principles. First, channels are traced starting from the center of a sphere, moving to adjacent points in a radially outward direction. Second, we identify the grid points with the largest channel width among the local points, which were not previously used as a sphere center, and use these as the centers of the spheres.

%
\begin{figure}[hbt]
    \centering
    \includegraphics[width=0.46\textwidth]{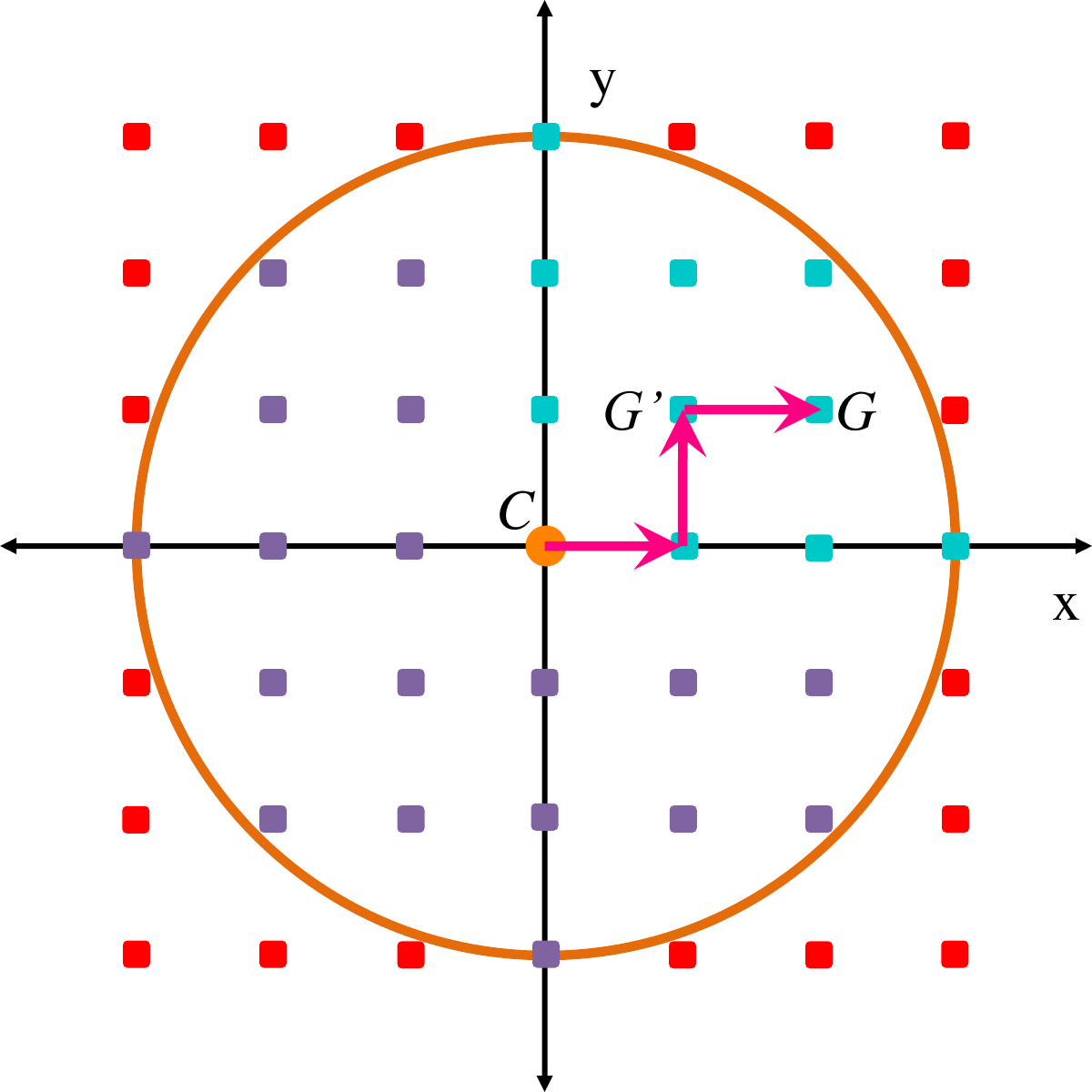}
    \caption{Two-dimensional visualization of tracing the grid points in a radially outward direction from the center of a pore. The center of the pore is denoted by $C$, and represented by an orange circle. The red squares represent the grid points outside of the pore. The purple and cyan colored squares represent the grid points inside the pores, and the grid points inside a quadrant that is currently being iterated over in the pore. The magenta colored line traces out a representative channel.}
    \label{fig:grid_iteration}
\end{figure}
%

The concept of tracing channels and assigning channel widths is illustrated in a two-dimensional diagram in Fig.~\ref{fig:grid_iteration}. This approach can be readily extended to three dimensions. The area within the circle is divided into quadrants, with channels traced one quadrant at a time. In Fig.~\ref{fig:grid_iteration}, the first quadrant is highlighted, where points are marked in cyan. Tracing a channel involves starting from the center $C$ and moving to adjacent points sequentially, adhering to allowed directions. In the first quadrant, movement is restricted to upward and rightward, reflecting the constraint of radially outward movement. Similar rules apply to other quadrants. For instance, in the third quadrant, only downward and leftward movements are permitted. This restriction significantly reduces the number of potential tracing, thus improving computational efficiency. Once all paths in the first quadrant are explored, the process proceeds to the second, third, and fourth quadrants. In three dimensions, the space is divided into octants, with three allowed movement directions within each octant.

The channel width at each grid point \( G \) is determined during the tracing process based on specific criteria. If \( G \) has not yet been assigned a channel width, there are two scenarios: (1) If the diameter of the fitted sphere centered at \( G \) exceeds the channel width of the preceding point \( G' \), the channel width of \( G \) is set to that of \( G' \). (2) Otherwise, \( G \)'s channel width is set equal to the diameter of the fitted sphere. If \( G \) has already been assigned a channel width, two cases can occur: (1) If the channel width of \( G' \) is greater than that of \( G \), \( G \) is updated to match \( G' \)'s channel width. (2) Otherwise, \( G \)'s channel width remains unchanged. Thus, all grid points within the circle can have their channel widths assigned based on the known width at the center, \( C \).

Once all the channels are traced inside a circle, the center of the circle $C$ is marked so that it will not be considered as a center in subsequent calculations. The algorithm ends once all the points have been marked. A list of all points encompassed by the circle is formed and sorted in a descending manner according to their channel width. This step is crucial for accurately determining the channel width through radially outward movement. The channel tracing process, as discussed earlier, is repeated for each point on the list, treating each as the new center \( C \). By segmenting the total area into circles, the algorithm reduces potential path lengths and variations, resulting in lower time and memory usage. It begins by initializing the list with grid points at the electrode-bulk electrolyte interface, assigning each point a channel width equal to the diameter of a fitted sphere centered at that point. Ultimately, all grid points receive a channel width that indicates the narrowest width an ion must traverse from the bulk electrolyte to reach that point.
%
\subsection{Miscellaneous}

The time-averaged parameters, such as coordination number, pore size, and channel width of adsorbed ions, were determined by averaging calculated profiles over 1000 frames collected during the 5 ns data-gathering simulation run. The coordination number of a counter-ion refers to the number of co-ions surrounding it. The occupied pore size was defined as the diameter of the largest sphere that could fit inside the electrode encompassing the counter-ion, without overlapping any carbon atoms. In the CDC models, the ring network was analyzed to identify individual rings and their constituting carbon atoms \cite{franzblau1991computation}. Carbon atoms shared by three rings were classified as basal plane atoms, while the remaining electrode atoms were categorized as edge atoms. The specific capacitance, $C_g$, was computed by \cite{pak2016molecular}
\begin{equation}
    \label{eq:cg}
    C_g = \frac{2Q_s}{m\Delta\psi} ,
\end{equation}
where $Q_s, m$ and $\Delta\psi$ are the steady-state charge, electrode mass, and potential difference, respectively.

\section{Results}
\label{result}

\subsection{Potential dependence of specific capacitance}

\begin{figure}[htb]
    \centering
    \includegraphics[width=0.46\textwidth]{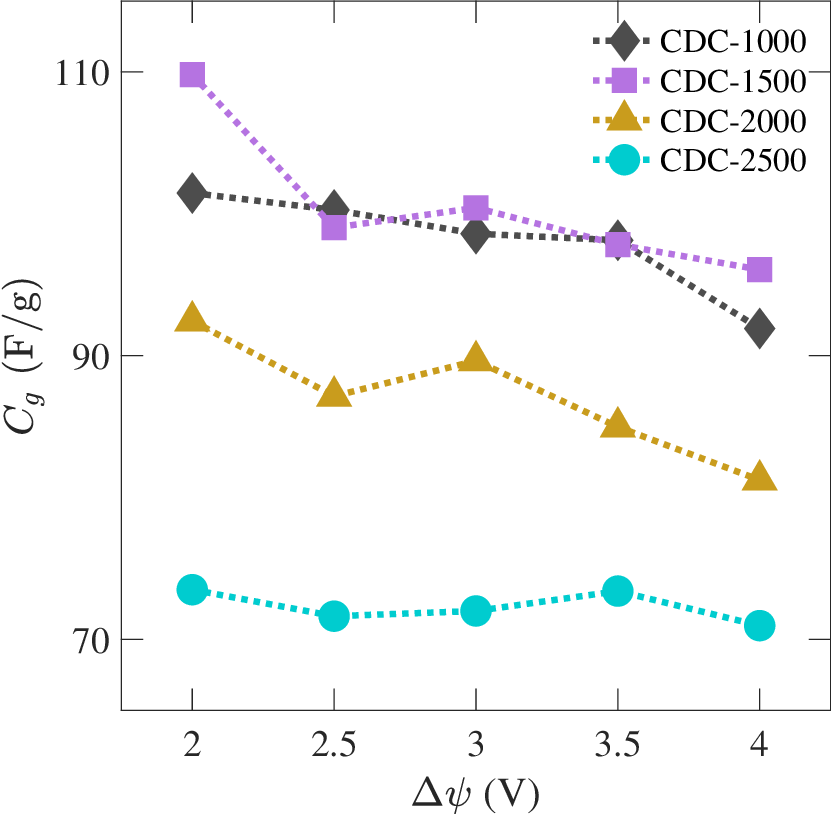}
    \caption{Specific capacitance ($C_g$) as a function of applied voltage ($\Delta\psi$) in supercapacitors with CDC-1000, CDC-1500, CDC-2000, and CDC-2500 electrodes and BMIM$^+$PF$_6^-$ electrolyte.}
    \label{fig:cg_v}
\end{figure}

The effect of electrode type and applied potential on $C_g$ is illustrated in Fig.~\ref{fig:cg_v}. The data points for different electrode types are clearly separated, indicating the significant influence of electrode type on $C_g$. The averaged $C_g$ values were 98.1, 100.6, 87.1, and 72.3 for CDC-1000, CDC-1500, CDC-2000, and CDC-2500, respectively. Strong qualitative agreement was found between simulated $C_g$ values and experimental data for SCs using CDC electrodes with tetraethylammonium [TEA$^+$] and tetrafluoroborate [BF$_4^-$] in acetonitrile electrolyte \cite{chmiola2006anomalous}. The experimental capacitances for devices with CDC electrodes synthesized at temperatures ranging from 500 to 1000 $^{\circ}$C were compared to those of electrode models generated with simulation temperatures of 1000, 1500, 2000, and 2500K. The experimentally observed $C_g$ values were 140.4, 143, 130, and 98.2 F/g for the synthesis temperatures of 500, 600, 800, and 1000 $^{\circ}$C, respectively. Although there were discrepancies in the exact $C_g$ values, both data sets exhibited similar percentage changes with temperature. These differences in values are attributed to the ion sizes in the electrolyte, as the experimental SCs feature TEA$^+$ and BF$_4^-$ ions of sizes 6.8 and 3.3 \AA, respectively. Conversely, the simulation utilized larger ions, BMIM$^+$ and PF$_6^-$, measuring 9.6 and 5.06 \AA, which contributed to the lower $C_g$ values observed in the simulations.

A decreasing trend in $C_g$ with increasing $\Delta\psi$ was observed in Fig.~\ref{fig:cg_v}. This trend indicates that the charging mechanism is primarily governed by ion adsorption, where counter-ions migrate to and adhere to the electrode under applied voltage. The polarization of electrode atoms compensates for the charge of the electrolyte ions. It was noted that at lower applied voltages, some pores became occupied by ions, limiting their capacity even as voltages increased. The pore size and channel width for accessing these pores significantly influenced the maximum number of ions they could hold. Consequently, the saturation of these pores contributed to the observed decline in $C_g$ with increasing voltage.

%
%

\subsection{Defining cutoff radius for sphere of influence}


Defining the sphere of influence for counter-ions is crucial for calculating DoC, counter charge, and other parameters that characterize the local ionic environment. Previously, the cutoff radius was set at the distance of the first minimum (6.3 \AA) in the partial RDF between counter-ions and electrode atoms \cite{merlet2013highly}. The partial RDF illustrates the local density variation of one type of particle relative to another. However, the position of the first minimum varies with the electrode, and in some cases, minima were absent, highlighting the need for a revised cutoff formulation \cite{liu2019carbons}. 

\begin{figure}[H]
    \centering
    \includegraphics[width=0.5\textwidth]{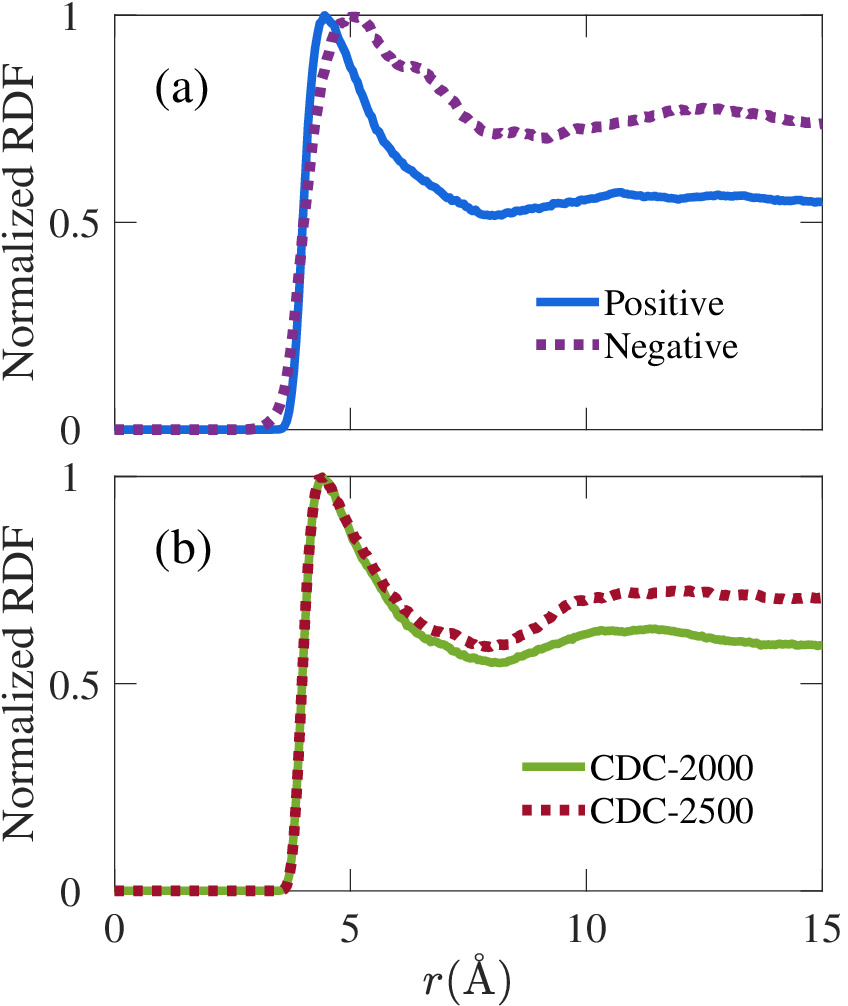}
    \caption{Normalized partial radial distribution function (RDF) calculated between counter-ion and carbon atom for counter-ions under an applied potential, $\Delta\psi = 3$ V. (a) The RDFs for CDC-1500 electrode do not have a clear minimum. (b) More defined minima are observed in the RDFs for CDC-2000 and CDC-2500 electrodes.} 
    \label{rdf_pn_1500_p_2000_2500}
\end{figure}

In our study, we calculated the partial RDF for CDC electrodes with BMIM$^+$PF$_6^-$ ions. The results for CDC-1500 electrodes, as shown in Fig.~\ref{rdf_pn_1500_p_2000_2500}(a),  indicate that the first minimum is poorly defined for both positive and negative electrodes, consistent with findings for CDC-1000. In contrast, CDC-2000 and CDC-2500 exhibited relatively well-defined minima at approximately 8.0 \AA,  as shown in Fig.~\ref{rdf_pn_1500_p_2000_2500}(b), deviating from the earlier cutoff of 6.3 \AA~\cite{merlet2013highly}. The atomistic models generated using the ABOP forcefield demonstrated better agreement with experimental data than those from previous studies~\cite{merlet2013highly, merlet2012molecular}.

However, the distance corresponding to the first minimum appeared to overestimate the sphere of influence \cite{sarkar2025influence}. In some cases, it includes carbon atoms that are shielded by layers closer to the ion, which should not fall within the sphere of influence, as the influence of the ion on them is significantly screened. This overestimation can lead to inaccuracies in the DoC value calculation. To address this, we developed a method in our prior work that accounts for overlapping atoms \cite{sarkar2025influence}, underscoring the necessity for an accurately defined sphere of influence in assessing the compensation charge on carbon atoms due to individual ions. 
%
\begin{figure}[hbt]
    \centering
    \includegraphics[width=0.46\textwidth]{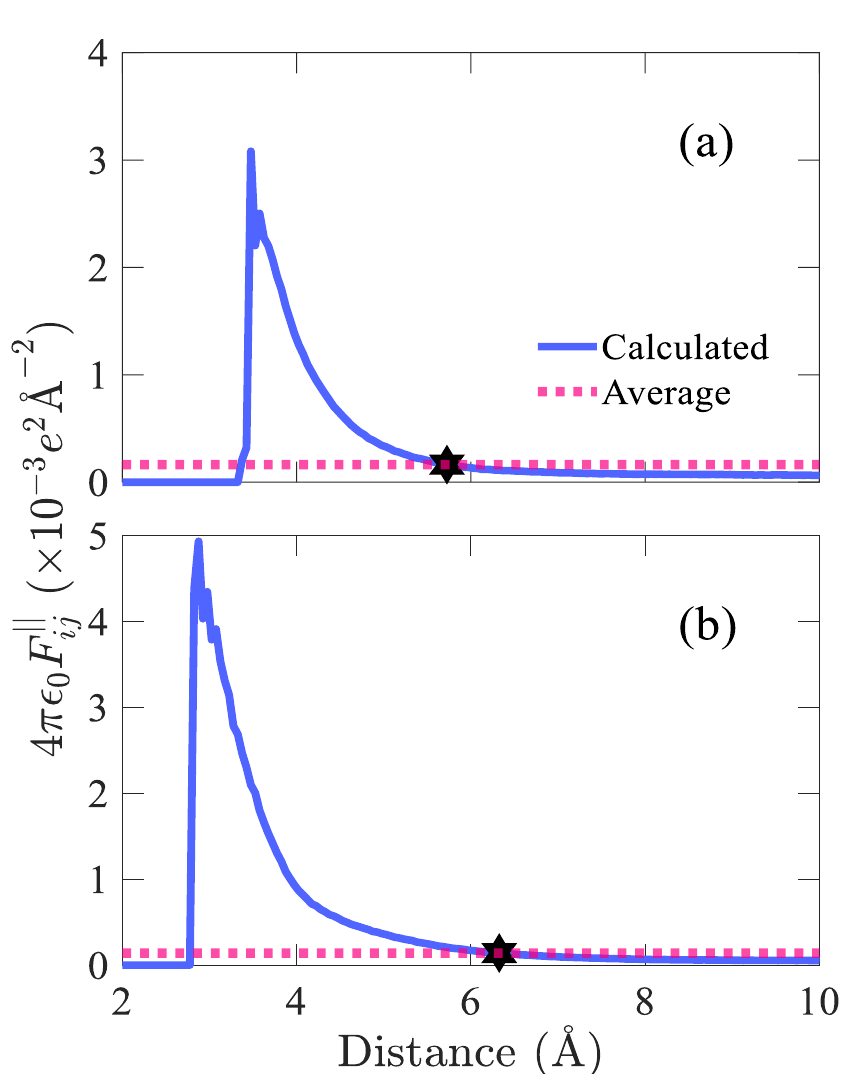}
    \caption{Variation of $F^{||}_{ij}$ with distance from counter-ions to electrode atoms in a supercapacitor device with CDC-1000 electrode type for (a) positive, and (b) negative electrodes under an applied potential, $\Delta\psi$ = 2V. The point defining the cutoff radius is highlighted by the black hexagram marker.}
    \label{fig:fij_avg}
\end{figure}
%

The rationale for defining the radius of influence based on the electric field profile stems from its impact on inducing charge on the surrounding carbon atoms due to individual adsorbed ions. The Coulomb force component profile was averaged over all adsorbed ions and analyzed to determine the spatial extent of significant ion influence. The Coulombic force component ($F^{||}_{ij}$), as defined by Eq.~(\ref{eq:force_comp}), averaged over all electrode atom counter-ion pairs and time at $\Delta\psi = 2V$ for the CDC-1000 electrode is illustrated in Fig.~\ref{fig:fij_avg}. The profile follows the expected inverse-squared trend with distance. Notably, the calculation of this profile only included positive $F^{||}_{ij}$ values as calculated from Eq.~(\ref{eq:force_comp}). The positive value indicated that the Coulomb force of the electrode atom had a scalar component towards the direction of the ion. Conversely, a negative value indicated that the force component of the electrode ion along the ion-atom line was pointing in the opposite direction to the ion.

To determine the cutoff radius from the $F^{||}_{ij}$ profile, the average value of the profile was calculated as denoted by the dotted magenta lines in Fig.~\ref{fig:fij_avg}. The distance at which the average value intersected with the decaying portion of the profile was considered as the cutoff radius, i.e., the radius of influence. The average was calculated considering the entire profile that extended up to 15 \AA. Our analysis showed that the average value saturated for profiles $\gtrsim12$ \AA~for all electrode types and polarities. As the Coulombic influence decays sharply, the electric field of most of the distant electrode atoms was not aligned in the direction of the adsorbed ions, leading to negative values of $F^{||}_{ij}$. As negative values of $F^{||}_{ij}$ were discarded from the calculation, the number of positive data points decreased with increasing distance, thus resulting in the saturated value of the average. 

The radii of influence for different electrode types and applied potentials are gathered in Table~\ref{tab:cutoff_radii}. The values corresponding to the negative electrode were larger than those of the positive electrode, due to the large size of the positive ion (BMIM$^+$) adsorbed in the negative electrode. The cutoff radii increased slightly with higher applied voltage.

\begin{table}
\centering 
\caption{The radius of the sphere of influence calculated from electric field analysis for different electrode types under the applied voltage($\Delta\psi$) of 2 and 4V.}
\resizebox{0.77\textwidth}{!}{
\begin{tabular}{c c c c c c}

\hline
\multirow{2}{*}{Potential (V)} & \multirow{2}{*}{Electrode polarity} & \multicolumn{4}{c}{Cutoff radius (\AA)} \\
\cline{3-6}

& & CDC-1000 & CDC-1500 & CDC-2000 & CDC-2500 \\
\hline

\multirow{2}{*}{2} & Positive & 5.72 & 5.77 & 5.67 & 5.82 \\

& Negative & 6.32 & 6.47 & 6.47 & 6.72 \\

\hline

\multirow{2}{*}{4} & Positive & 5.87 & 6.0 & 5.97 & 6.17 \\

& Negative  & 6.37 & 6.62 & 6.87 & 6.67 \\

\hline

\end{tabular}
}
\label{tab:cutoff_radii}
\end{table}

%
\subsection{Edge and basal plane carbon atoms}

\begin{figure}[htb]
    \centering
    \includegraphics[width=0.46\textwidth]{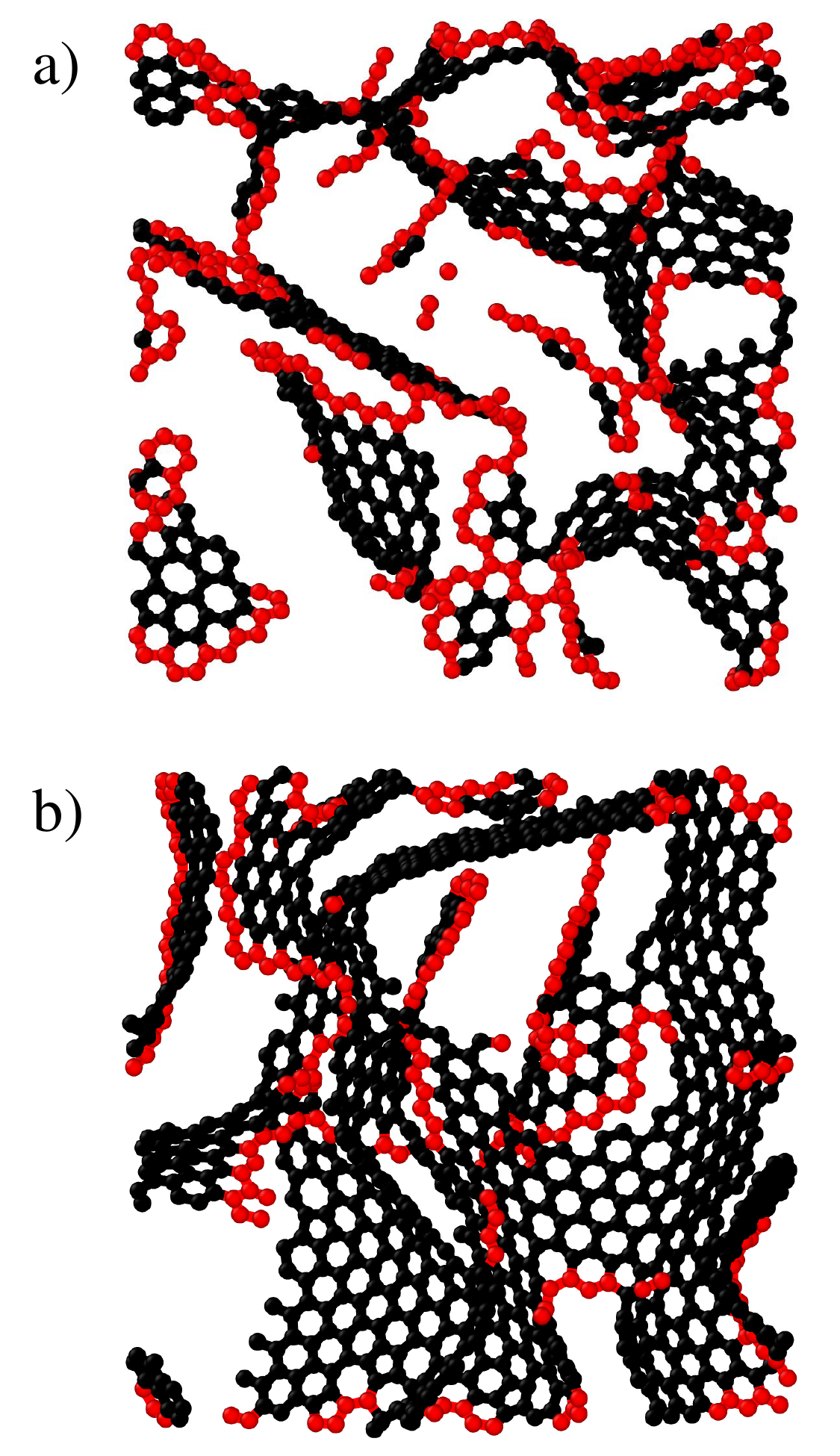}
    \caption{Cross-sectional view along $yz$ plane of 10 \AA~slab of (a) CDC-1500, and (b) CDC-2000 electrodes, respectively, highlighting the edge and basal plane carbons by red and black spheres, respectively.}
    \label{fig:sch_edge_plane}
\end{figure}

In our previous study, electrolyte ions were found to adsorb at a minimal distance from edge atoms, resulting in significant charge accumulation on carbon atoms \cite{sarkar2025influence}. This observation prompted a detailed investigation into the charge distribution between edge and basal plane atoms. Figure \ref{fig:sch_edge_plane} visually illustrates the arrangement of these atoms within the electrode structures, which primarily consist of graphitic sheets featuring hexagonal rings, with edges terminating at edge atoms. 

\begin{table} [H]
\centering 
\caption{Number of edge and basal plane site carbon atoms in different electrodes.}
\resizebox{0.6\textwidth}{!}{
\begin{tabular}{c c c c c}

\hline

& \multicolumn{4}{c}{Number of atoms} \\

\cline{2-5}

& CDC-1000 & CDC-1500 & CDC-2000 & CDC-2500 \\
\hline

Edge & 2414 & 1699 & 735 & 541 \\

Basal plane & 1601 & 2313 & 3289 & 3461 \\

Total & 4015 & 4012 & 4024 & 4002 \\

\hline

\end{tabular}
}
\label{tab:site_count}
\end{table}

As detailed in Table \ref{tab:site_count}, the number of edge atoms decreased with increasing synthesis temperature of the CDC material. Notably, the CDC-1000 electrode exhibited the highest proportion of carbon atoms (60\%) at the edge sites, while the CDC-2500 electrode displayed the lowest. Since the total atom count remained relatively constant across all electrodes, a greater number of edge atoms indicated that the carbon structure comprised smaller graphitic sheets compared to the others. This trend aligns with previous findings that higher synthesis temperatures result in the formation of larger, more planar graphitic sheets within the CDC models \cite{sarkar2024structurally}.

%

The ion penetration and pore filling were influenced by the applied voltage and electrode type, as outlined previously. The relationship between edge and basal plane carbon atoms affected by counter-ions is summarized in Table \ref{tab:counter_carbon_count}. This count was derived from identifying carbon atoms within the sphere of influence of the counter-ions across all frames from the last 1000 frames of the simulation. The ratio of influenced edge to basal plane carbon atoms mirrored the overall ratio of these atom types for both CDC-1000 and CDC-1500. However, CDC-2000 and CDC-2500 did not exhibit this same pattern. In these electrodes, edge atoms constituted a smaller percentage ($<$20\%) compared to their lower-temperature counterparts ($>$40\%). This limited accessibility of edge atoms to ions entering through the positive (upper) and negative (lower) electrode surfaces contributed to deviations in the ratios of total versus influenced electrode atoms. 

\begin{table}
\centering 
\caption{Number of edge and basal plane atoms which were inside the sphere of influence of the counter-ion during the data-gathering simulation run.}
\resizebox{1\textwidth}{!}{
\begin{tabular}{c c c c c c c}

\hline
\multirow{2}{*}{Potential (V)} & \multirow{2}{*}{Electrode polarity} & \multirow{2}{*}{Site} & \multicolumn{4}{c}{Number of atoms} \\
\cline{4-7}

& & & CDC-1000 & CDC-1500 & CDC-2000 & CDC-2500 \\
\hline

\multirow{4}{*}{2} & \multirow{2}{*}{Positive} & Edge & 1765 & 1290 & 569 & 253 \\

& & Basal plane & 1283 & 1999 & 2910 & 2635 \\

& \multirow{2}{*}{Negative} & Edge & 1878 & 1341 & 510 & 238 \\

& & Basal plane & 1328 & 2047 & 2727 & 2512 \\

\hline

\multirow{4}{*}{4} & \multirow{2}{*}{Positive} & Edge & 1961 & 1393 & 599 & 256 \\

& & Basal plane & 1374 & 2123 & 2957 & 2703 \\

& \multirow{2}{*}{Negative} & Edge & 1937 & 1410 & 544 & 266 \\

& & Basal plane & 1368 & 2112 & 2840 & 2708 \\

\hline

\end{tabular}
}
\label{tab:counter_carbon_count}
\end{table}

\begin{figure}[htbp]
    \centering
    \includegraphics[width=1.0\textwidth]{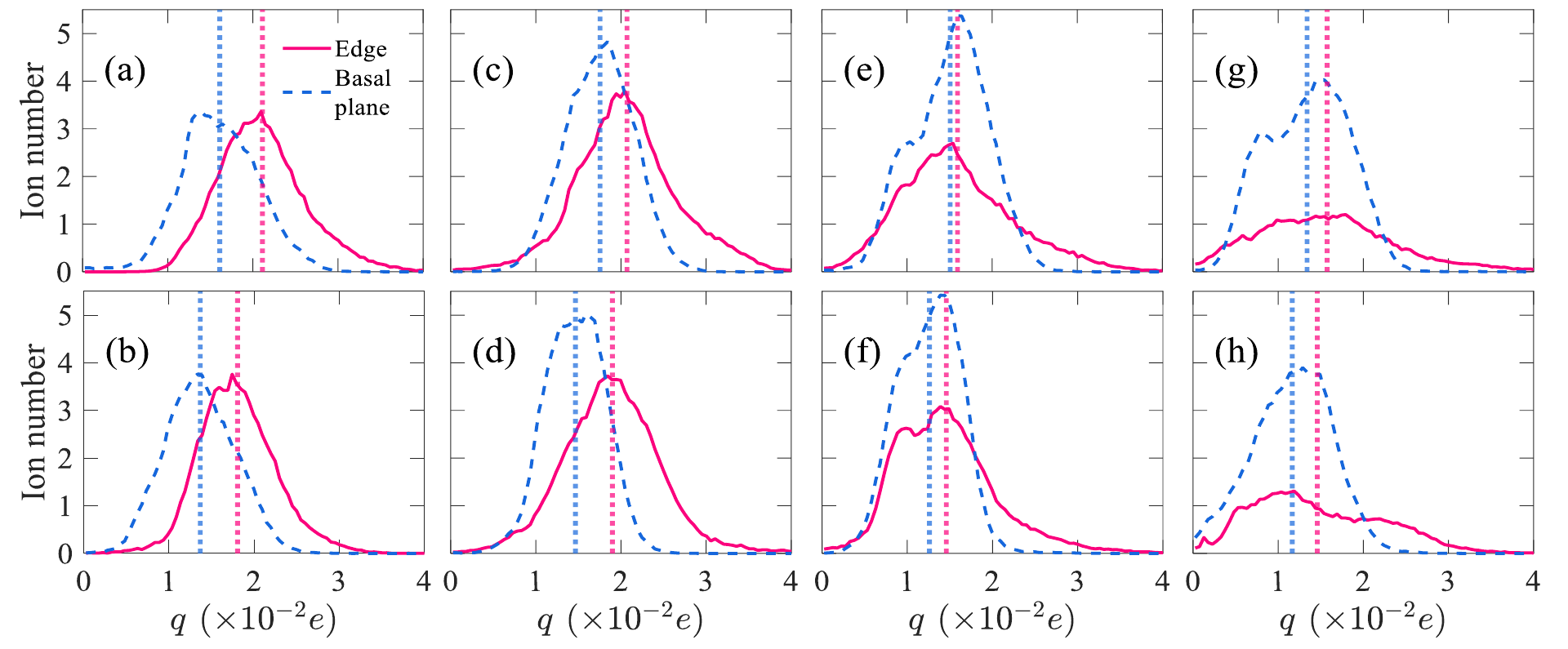}
    \caption{Average counter-ion number variation with charge per atom for edge and basal plane carbon atoms in (a-b) CDC-1000, (c-d) CDC-1500, (e-f) CDC-2000, and (g-h) CDC-2500 electrodes under an applied potential of 2V. The top rows (a, c, e, f) and bottom rows (b, d, f, h) correspond to the positive and negative electrodes, respectively. The average charge of a profile is denoted by the vertical dotted line of the same color.}
    \label{fig:edge_plane_peratom}
\end{figure}

The influence of edge and basal plane sites on charge accumulation is examined by calculating the charge per atom from electrode atoms across different sites. For each counter-ion, the total charge associated with edge and basal plane atoms was assessed, and the charge per atom for various sites was determined. Subsequently, the ion counts were binned according to the charge per atom, as illustrated in Fig.~\ref{fig:edge_plane_peratom}. The profiles of edge atoms exhibited a rightward shift in both positive and negative electrodes across all electrode types. The average charge values, denoted by the vertical lines, further highlight the significant difference in charge per atom between edge and basal plane atoms. Notably, the profile values decreased along the $y$ axis from CDC-1000 to CDC-2500 electrodes, due to the reduction in the number of edge atoms. This decrease leads to fewer counter-ions adsorbed near edge sites as the electrode synthesis temperature increases. Therefore, it is clear that edge atoms possess a higher average charge compared to basal plane atoms, implying that electrodes with a greater number of edge atoms can store more charge.

%

The conclusions drawn from our analysis of edge and basal plane atoms are consistent with recent experimental findings \cite{liu2024structural}, which reveal that a disordered structure characterized by smaller graphitic domains is a crucial factor contributing to enhanced capacitance. This experimental study challenges the longstanding notion that pore size is the primary determinant of capacitance. Specifically, reference \cite{liu2024structural} investigated the chemical shifts observed in nuclear magnetic resonance spectra for both a free electrolyte and carbon saturated with electrolyte. The resulting difference between these two chemical shifts was designated as $\Delta\delta$. A higher value of $\Delta\delta$ indicates greater structural order, while a lower value implies increased disorder. Through the examination of 20 different SCs, the study found no significant correlation between capacitance and average pore size or surface area. However, a striking correlation emerged between capacitance and $\Delta\delta$ values, indicating that carbons with smaller ordered domains are associated with enhanced capacitance. The authors of reference \cite{liu2024structural} proposed that the edge atoms present in such disordered structures may facilitate a higher accommodation of charge. Our findings bolster this hypothesis, showing that carbons with smaller graphitic sheets indeed possess a greater number of edge atoms, which contribute to an increased charge density in comparison to basal plane atoms.

%
\subsection{Coordination number and channel width}

\begin{figure}[htb]
    \centering
    \includegraphics[width=1\textwidth]{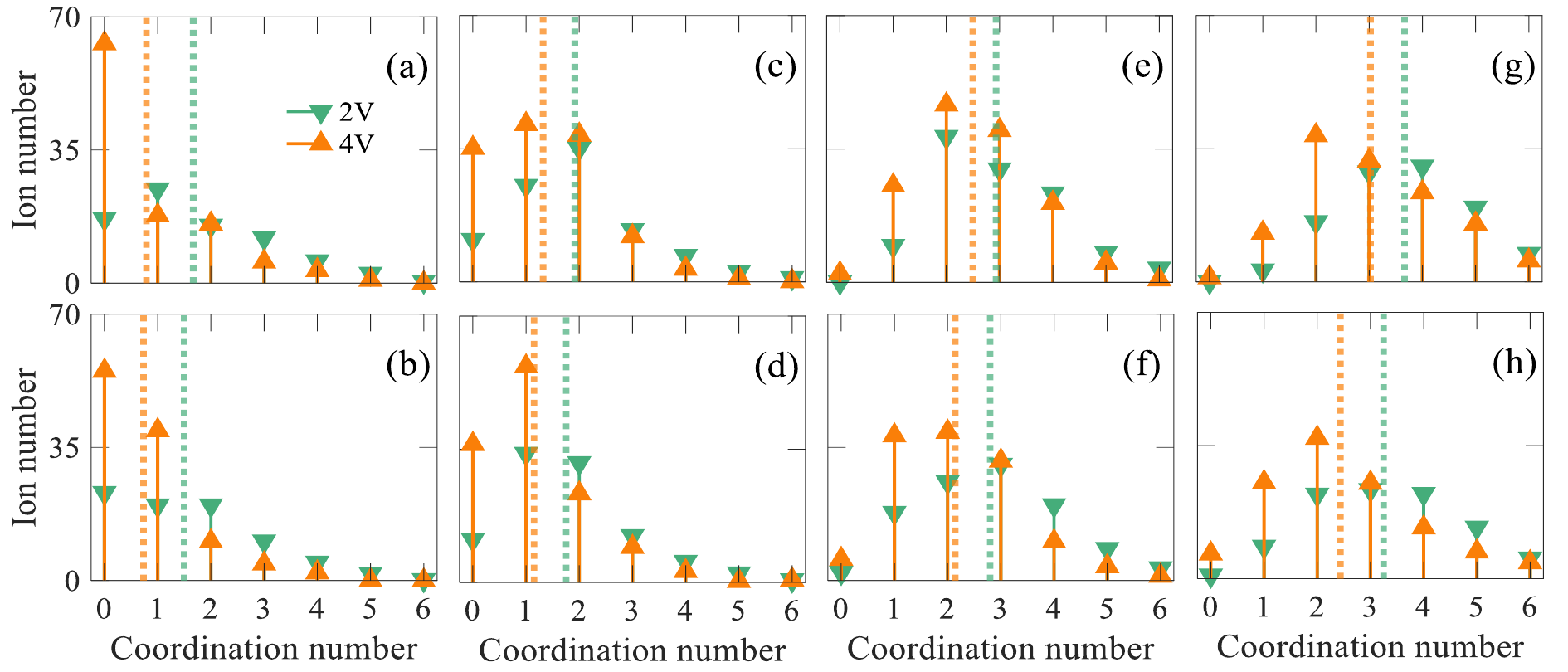}
    \caption{Coordination number distribution of counter-ions in (a, c, e, f) positive, and (b, d, f, h) negative electrodes for (a-b) CDC-1000, (c-d) CDC-1500, (e-f) CDC-2000, and (g-h) CDC-2500 electrode types, respectively, under applied potentials of 2 and 4V. The vertical dotted lines represent the average coordination number associated with the profile of the same color.}
    \label{fig:co_dist}
\end{figure}
%

The effect of applied voltage on counter-ion decoordination is illustrated in Fig.~\ref{fig:co_dist}. In the bulk electrolyte, the average coordination number for both anion and cation was 6. However, all electrode types exhibited average coordination numbers well below this value, with the lowest recorded at 0.74 for the negative CDC-1000 electrode. Notably, decoordination increased with rising applied voltage, as evidenced by the decline in average coordination numbers from 2V to 4V for all electrodes. This trend highlights the strong correlation between the total induced charge and the coordination number of counter-ions, attributable to the reduced screening effect from co-ions. At an applied potential of 4V, the average coordination numbers were 0.78, 1.32, 2.5, and 3.0 for the positive CDC-1000, CDC-1500, CDC-2000, and CDC-2500 electrodes, respectively. This result demonstrates that the type of electrode significantly influences the decrease in average coordination numbers, reflecting a pronounced leftward shift in the coordination number profile.

\begin{figure}[H]
    \centering
    \includegraphics[width=0.55\textwidth]{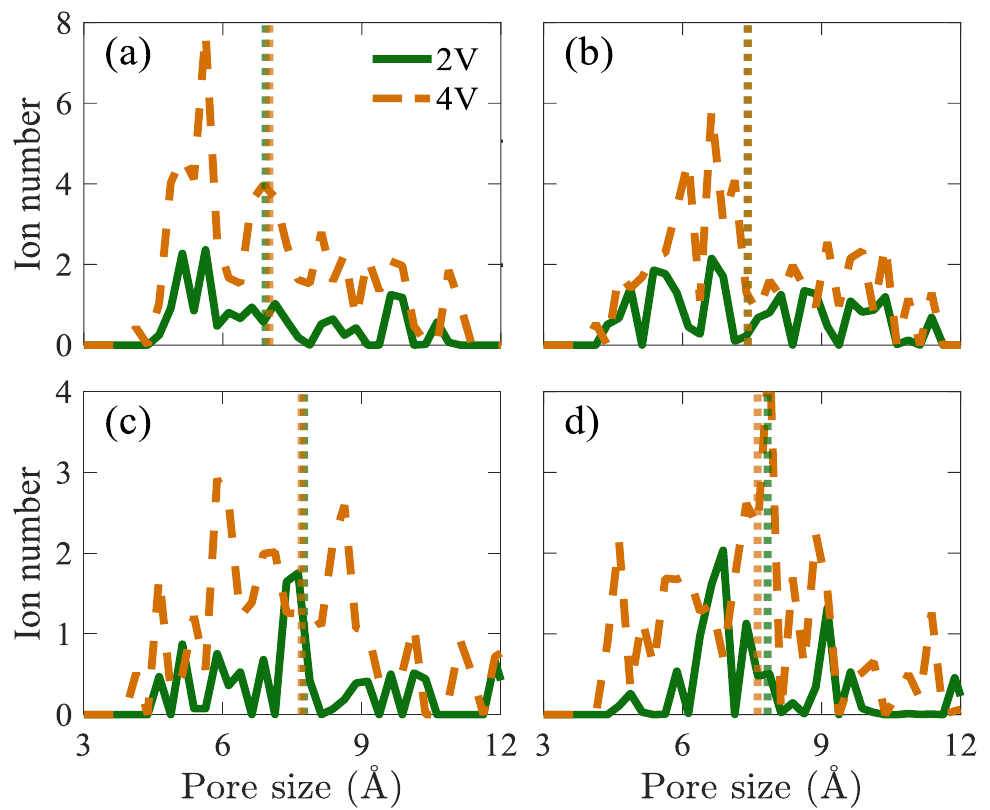}
    \caption{Occupied pore size distribution calculated from fitting the largest sphere not overlapping with any carbon atom, and encompassing the counter-ion, for (a-b) CDC-1000, and (c-d) CDC-1500 electrodes with (a,c) positive and (b,d) negative polarities. The vertical dotted lines correspond to the average values of the profiles with the same color.}
    \label{fig:occ_pore_co_0}
\end{figure}

A key observation from Fig.~\ref{fig:co_dist} is the minimal occurrence of ions with a coordination number of zero, even at the maximum applied voltage of 4V, for the CDC-2000 and CDC-2500 electrodes, in contrast to their lower temperature counterparts. The left-shifted coordination number profile in electrodes synthesized at lower temperatures has been attributed to smaller pore sizes~\cite{sarkar2025influence}. However, further analysis indicates that the size of the pores occupied by counter-ions is not the primary determinant of coordination number. Figure \ref{fig:occ_pore_co_0} illustrates the distributions of pore size for counter-ions with a coordination number of zero in the CDC-1000 and CDC-1500 electrodes only, as the CDC-2000 and CDC-2500 electrodes contained too few such ions for reliable statistical analysis. The occupied pore sizes for highly decoordinated counter-ions ranged from 3.75 to 12 \AA. Notably, the highest peaks of these distributions were all below 8 \AA, with most ions found in pores smaller than this size. Nevertheless, the significant presence of highly decoordinated counter-ions in larger pores challenges the assumption that smaller pore sizes are the main factor in isolating counter-ions.

%

\begin{figure}[htb]
    \centering
    \includegraphics[width=0.55\textwidth]{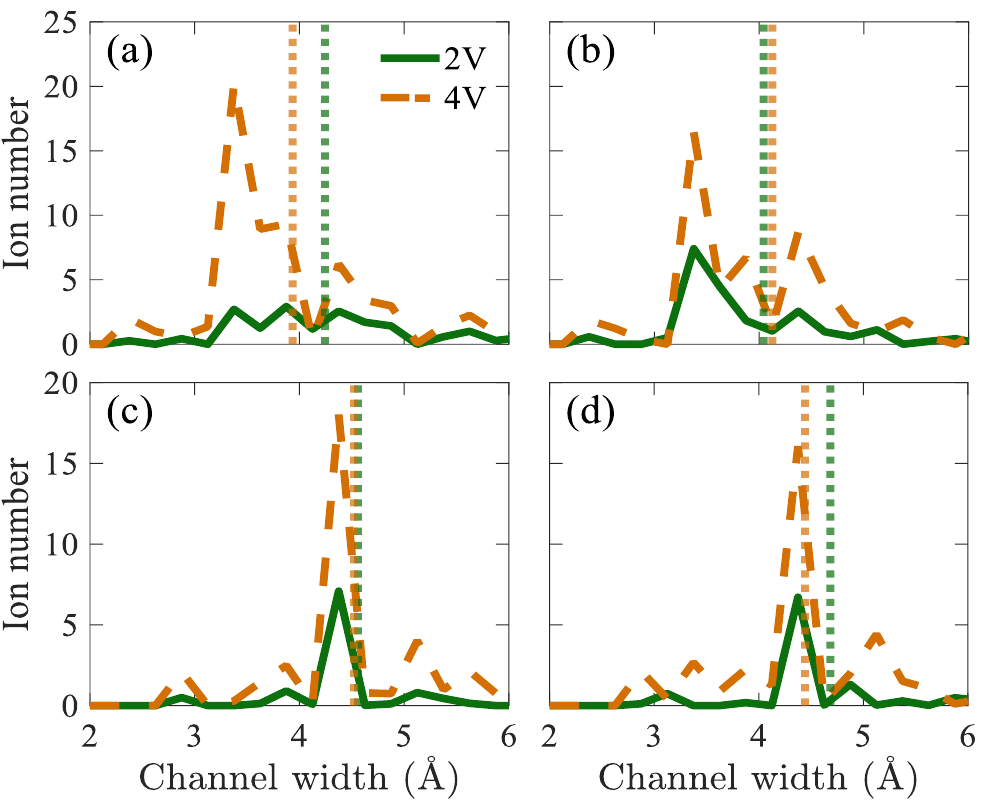}
    \caption{Channel width distribution, \textit{i.e.}, the diameter of the smallest sphere a counter-ion had to pass through to reach the pore for (a-b) CDC-1000, and (c-d) CDC-1500 electrodes with (a,c) positive and (b,d) negative polarities. The vertical lines correspond to the average values of the profiles with the same color.}
    \label{fig:cw_co_0}
\end{figure}

The distribution of channel widths for counter-ions with a coordination number of zero displayed a narrow range, as illustrated in Fig.~\ref{fig:cw_co_0}. These profiles were characterized by a single prominent peak, contrasting with the broader distribution observed for occupied pore sizes, which featured multiple peaks. The analysis indicated that the isolated and decoordinated counter-ions were preferentially adsorbed through narrow channels, independent of the pore size. Both CDC-1000 and CDC-1500 exhibited similar average channel widths of approximately 4 \AA. Notably, over 80\% of the counter-ions were found in pores with channel widths of 4.675 \AA~or less, highlighting that narrow channel widths are essential for the adsorption of counter-ions with a coordination number of zero.  

%

\begin{figure}[H]
    \centering
    \includegraphics[width=1\textwidth]{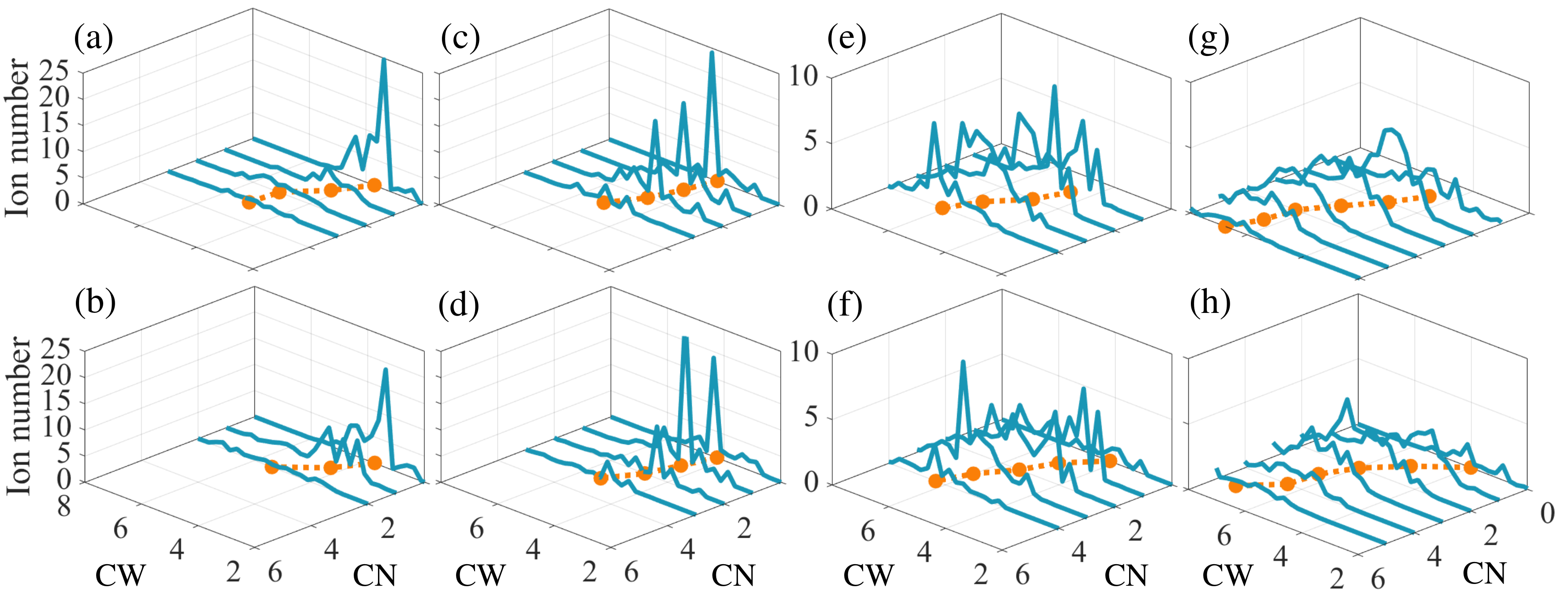}
    \caption{Distribution of ion number as a function of channel width (CW) and coordination number (CN) for (a-b) CDC-1000, (c-d) CDC-1500, (e-f) CDC-2000, and (g-h) CDC-2500 electrodes with the potential difference $\Delta\psi = 4$V. The (a,c,e,g) top, and (b,d, f, h) bottom rows of plots correspond to positive and negative electrodes, respectively. The average channel width corresponding with each coordination number is presented by the orange markers.}
    \label{fig:3D_cw_co}
\end{figure}

The distribution of channel width for counter-ions with varying coordination numbers is illustrated in Fig.~\ref{fig:3D_cw_co}. For statistical significance, a coordination number's corresponding channel width distribution is plotted if the number of data points associated with the distribution exceeds 5\% of the total number of data points. For both polarities of the electrodes, the average channel width consistently increased with higher coordination numbers, suggesting that greater decoordination necessitates narrower channel widths. This inverse relationship between coordination number and channel width is further illustrated by the observation that the highest peak in the distribution appears at a greater channel width as coordination number increases. The average channel widths for the coordination number of 1 were similar across different electrodes, measuring 4.24, 4.37, 4.56, and 4.54 \AA~for positive CDC-1000, CDC-1500, CDC-2000, and CDC-2500 electrodes, respectively. Similarly, the average channel widths for coordination number 2 in positive electrodes ranged from 4.64 to 5 \AA. The similar values of the averages indicated that the ions achieved a similar degree of decoordination as they passed through similar channel widths, irrespective of the electrode type.

%

\section{Conclusion}
\label{conslusion}
The SCs were modeled using an atomistic representation of CDC material as the electrode, incorporating four distinct PSDs and average pore sizes. The applied voltage range of 2 to 4 V revealed a decreasing trend in specific capacitance: lower voltages enabled better filling of pores with electrolyte ions, while higher voltages constrained ion accommodation, resulting in reduced capacitance. Edge electrode atoms---those shared by two or fewer rings---were analyzed to gauge disorder within the electrode structure. By maintaining consistent density across electrodes, a greater number of edge atoms indicated smaller graphitic sheets and increased disorder. This disorder exhibited an inverse relationship with synthesis temperature, with the CDC-1000 electrode showing the highest disorder (approximately 60\% edge atoms). Interestingly, the presence of edge atoms allowed disordered structures to carry more charge than their organized counterparts.

The channel width of a pore, representing the narrowest passage for ion movement from the bulk electrolyte, was critical for ion decoordination. Analysis of pore size and channel width distribution of counter-ions demonstrated that channel width had a more significant impact on decoordination than pore size. The trend of increasing average channel width with rising coordination number reinforced this notion. Overall, this study linked device performance metrics to structural features such as edge atom concentration and pore channel width, providing valuable insights for understanding capacitance variation and optimizing electrodes with diverse porous architectures.

\section*{Data availability}
\noindent All data in the paper are present in the main text, which will also be available from the corresponding author upon reasonable request.
\section*{Conflicts of interest}
\noindent There are no conflicts of interest to declare.

%



 \bibliographystyle{elsarticle-num} 
 \bibliography{biblio_}





\end{document}